\documentclass[conference]{IEEEtran}

\usepackage[utf8]{inputenc}
\usepackage[dvips]{graphicx}
\usepackage{amsmath}
\usepackage[psamsfonts]{amssymb} 
\usepackage{stmaryrd}
\usepackage[linesnumbered,ruled,lined,titlenotnumbered]{algorithm2e}
\input xy
\xyoption{all}

\newtheorem{definition}{Definition}
\newtheorem{lemma}{Lemma}

\newtheorem{corollary}{Corollary}

\newcommand{\nodeset}[0]{{\cal N}}
\newcommand{\contactset}[0]{{\cal C}}
\newcommand{\distset}[0]{{\cal D}}

\newcommand{\schedule}{\oslash}

\begin{document}

\title{Predictable Disruption Tolerant Networks \\
  and Delivery Guarantees}

\author{Jean-Marc Fran\c{c}ois and Guy Leduc \\
Research Unit in Networking (RUN)\\
  Universit\'e de Li\`ege, Belgium\\
  {\{francois,leduc\}@run.montefiore.ulg.ac.be}}

\maketitle
\thispagestyle{empty}

\begin{abstract}
This article\footnote{This work has been partially supported by the
 Belgian Science Policy in the framework of the IAP program (Motion
 P5/11 project) and by the European E-Next NoE and IST-FET ANA
 project.}
 studies disruption tolerant networks (DTNs) where each
node knows the probabilistic distribution of contacts with other
nodes. It proposes a framework that allows one to formalize the
behaviour of such a network.  It generalizes extreme cases that have
been studied before where {\em (a)} either nodes only know their
contact frequency with each other or {\em (b)} they have a perfect
knowledge of who meets who and when.  This paper then gives an example
of how this framework can be used; it shows how one can find a packet
forwarding algorithm optimized to meet the 'delay/bandwidth
consumption' trade-off: packets are duplicated so as to
(statistically) guarantee a given delay or delivery probability, but
not too much so as to reduce the bandwidth, energy, and memory
consumption.
\end{abstract}

%
%
%
\section{Introduction}
\label{sec-introduction}
%

{\em Disruption} (or {\em Delay}) {\em Tolerant Networks} (DTNs,
\cite{Z06}) have been the subject of much research activity in the
last few years, pushing further the concept of Ad~Hoc networks.  Like
Ad~Hoc networks, DTNs are infrastructureless, thus the packets are
relayed from one node to the next until they reach their
destination. Moreover, in DTNs node clusters can be completely
disconnected from the rest of the network.  In this case, nodes must
buffer the packets and wait until node mobility changes the network's
topology, allowing the packets to be finally delivered.

A network of Bluetooth-enabled PDAs, a village intermittently connected
{\em via} low Earth orbiting satellites, or even an interplanetary
Internet (\cite{BHTFXDSW03}) are examples of disruption tolerant networks.


The atomic data unit is a group of packets to be
delivered together. In DTN parlance, it is called a {\em message} or a 
{\em bundle}; we use the latter in the following.

Routing in such networks is particularly challenging since it requires
to take into account the uncertainty of mobiles movements.  The first
methods that have been proposed in the literature are pretty radical
and propose to forward bundles in an ``epidemic'' way
(\cite{VB00,SH03,JOWMPR02}), {\em i.e.}, to copy them each time a new
node is encountered.  This method of course results in optimum delays
and delivery probabilities, at the expense of an extremely high
consumption of bandwidth (and, thus, energy) and memory.  To mitigate
those shortcomings, the epidemic routing has been enhanced using
heuristics that allow the propagation of bundles to a subset of all
the nodes (\cite{TR01,SPR04,SPR05}).

Since node's buffer memory is not unlimited, a cache mechanism has
been proposed, where the most interesting bundles are kept ({\em i.e.}
those that are likely to reach their destination soon) and the others
are discarded when the cache is full
(\cite{NPB03,WJMF05,LDS03,TZZ03,JLW05,LFC05}). Those schemes
must thus guess when a bundle will reach its destination, which is
most of the time computed thanks to frequency contact estimation
(which reflects the probability that two given nodes meet in the
future).

Few papers explore how the expected delay could be more precisely
estimated (notable exceptions are \cite{SBRJ02,MHM05}). It has been
proved (\cite{M04}) that a perfect knowledge of the future node
meetings allows the computation of an optimal bundle routing.

This short overview emphasizes two shortcomings:
\begin{itemize}
\item Certain networks might be highly predictable ({\em e.g.} nodes
  are satellites and links appear and vanish as they revolve around
  their planet), others are much more chaotic.  Previous work suppose
  either that nodes contacts are perfectly deterministic and known in
  advance, or that only the contact frequency is known for each pair
  of nodes.  We propose to generalise these approaches and suppose
  that each node knows a probability distribution of contacts in the
  (near) future.
\item \cite{JOWMPR02} underlines the tradeoff between bundle delivery
  guarantees and bandwidth/energy consumption: copying the bundles is
  costly since, in mobile networks, those resources are both scarce.
  Current schemes use a cache mechanism that ensures each node only
  receives the most relevant bundles, which somehow mitigates this
  problem, but does not provide any rationale, except the need to cope
  with mobiles limited memory.  We propose to route the bundles
  according to the delivery or delay guarantees required by the user,
  thus only duplicating packets when it is beneficial.
\end{itemize}

This paper is organised as follows.  Section~\ref{sec-contacts}
presents a way to model the contacts between the nodes of a
predictable network.  Sections~\ref{sec-contacts} and
\ref{sec-operators} show how the end-to-end delay of bundles can be
predicted.  Sections~\ref{sec-guarantees} and \ref{sec-algo} give a
routing algorithm that allows to deliver bundles in a manner that
meets a given guarantee.  Section~\ref{sec-conclusions} concludes.

\begin{figure*}[h!bt]
\begin{center}\includegraphics[scale=.95]{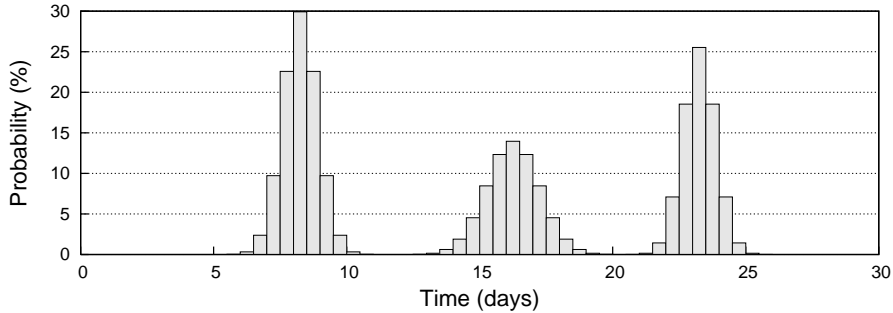}\end{center}
\caption{\textbf{Contact profile of a node pair over a month:
  example.}  The height of a bar gives the probability that two nodes
  meet ({\em at least} once) during the corresponding 12-hour time
  period.  Here, nodes are supposed to meet at the beginning of each
  week, but the exact day is unknown.}
\label{fig-contacts}
\end{figure*}

\begin{figure*}[h!bt]
\begin{center}\includegraphics[scale=.95]{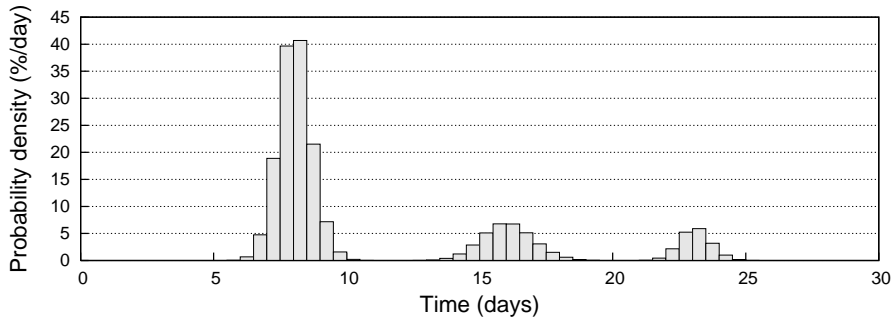}\end{center}
\caption{\textbf{First contact probability distribution}
  corresponding to the contact profile figure~\ref{fig-contacts}.
  (Each bar corresponds to a 12-hour period.)}
\label{fig-distrib0}
\end{figure*}

%
\section{Predictable future contacts}
\label{sec-contacts}
%

The network is composed of a finite set of wireless nodes $\nodeset$
that can move and thus, from time to time, come into contact.

In the sequel, a {\em contact} between two nodes happen when those
nodes have setup a bi-directional wireless link between them.  A
contact is always considered long enough to allow all the required
data exchanges to take place\footnote{This is a major difference with
\cite{M04} which does not neglect bundle transmission times.}.

\subsection{Contact profiles}

We expect the mobiles motion to be, to a certain extent, predictable,
yet obviously the degree of predictability varies from one network to
another.  Sometimes nodes motion is known in advance because they must
stick to a given schedule ({\em e.g.} a network of buses) or because
their trajectory can easily be modelled ({\em e.g.} nodes embedded in
a satellite).  Other networks are less predictable, yet not totally
random: colleagues could be pretty sure to meet every day during
working hours, without any other time guarantee.  Mobile nodes
behaviour could also be learnt automatically so as to extract cyclical
contact patterns.

We therefore suppose that each node pair ${\{a,b\} \subset \nodeset}$
can estimate its contact probability for each time step in the near
future. We call it a {\em contact profile} and denote it $C_{ab}:
\mathbb{N} \rightarrow [0,1]$.  The time step duration should be
chosen small compared to the expected network's end-to-end delay.
Figure~\ref{fig-contacts} gives an hypothetical contact profile.  In
the following, we suppose the profile known for each node pair.

Contact profiles can easily represent situations usually depicted
in the literature:
\begin{itemize}
\item A constant profile $C_{ab}(t)=k$ describes a node pair that only
  knows its contact frequency.  For example, the profile
  ${C_{ab}(t)=1/30}$ (probability of contact per day) corresponds to
  two nodes $a$ and $b$ meeting once a month on average.
\item Perfect knowledge of nodes meeting times results in a profile
  made of peaks: $\forall t\in\mathbb{N}: C_{ab}(t)\in\{0,1\}$.
\end{itemize}
In practice, unknown contact profiles could be replaced by a null
function to get a defensive approximation of their behaviour.

The following sections aim at studying how bundles propagate from one
node to another in a network whose nodes' contact profiles are known.

\begin{figure*}[htb]
\begin{center}\includegraphics[scale=.95]{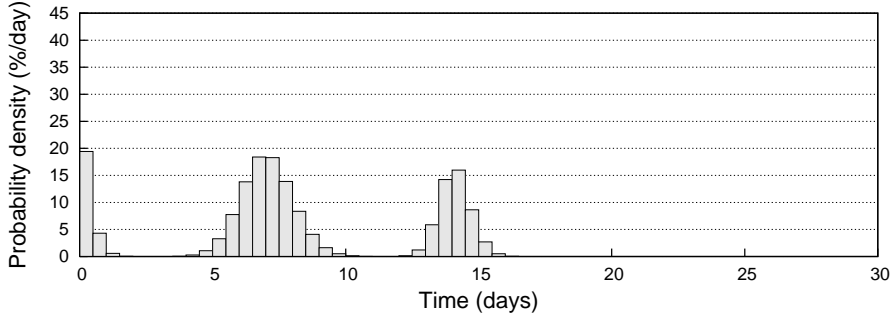}\end{center}
\caption{\textbf{The contact probability density
   \boldmath$D_{ab}(9,\cdot)$} matching the contact profile given in
   figure~\ref{fig-contacts}.}
\label{fig-distrib9}
\end{figure*}

\subsection{First contact distribution}

It is easy to deduce the probability distribution of a (first) contact
at time $t$ between nodes $a$ and $b\in\nodeset$ given their
profile $C_{ab}$; we denote this distribution $d_{ab}$.  Since the
probability of a first contact at time $t$ is the probability of
meeting at time step $t$ times the probability not to meet at time
steps $0,1,\dots,t-1$, we have:
\begin{equation}
d_{ab}(t) = C_{ab}(t)\; \prod_{i=0}^{t-1} \bigl(1-C_{ab}(i)\bigr) \qquad
  \forall a,b\in \nodeset, \;\forall t\in\mathbb{N}
\end{equation}

The distributions domain is $\mathbb{N}$ since contact profiles have
been defined using discrete time steps.  We extend the distributions
to $\mathbb{R}$ to get rid of this artifact.  Notice that $d_{ab}$ is
not a well-defined probability distribution since its integral over
its domain is not equal to $1$: two nodes might never meet.  Those
considerations directly lead to the definition of the first contact
distribution set.

\begin{definition}
\label{def-contactset}
The first contact distribution set, $\contactset$, is the set of
functions\footnote{$\mathbb{R^+}$ denotes the set of positive reals.}
\mbox{$f: \mathbb{R^+} \rightarrow \mathbb{R^+}$} such that
\mbox{$\int_0^\infty f(x) \,dx \leq 1$}.
\end{definition}

Contact profiles have a shortcoming: they do not allow us to express
contact interdependencies; for example, they cannot model that two
nodes are certain to meet during the weekend without knowing exactly
which day.  First contact distributions have no such limitations.
Therefore, when it is possible, one could find preferable to generate
them directly without relying on contact profiles.

Figure~\ref{fig-distrib0} gives the $d_{ab}$ distribution
corresponding to the contact profile $C_{ab}$ depicted in
figure~\ref{fig-contacts}.


Notice that if a bundle is delivered directly from $a$ to $b$, knowing
the first contact distribution allows an easy verification of a large
spectrum of guarantees, such as the average delay or the probability
of delivery before a certain date.

%
\section{Delivery distributions}
\label{sec-distributions}
%

\subsection{Definition}

First contact distributions can be generalized to take into account
the knowledge that no contact were made before a certain date.

Let $D_{ab}(T,t)$ be the probability distribution that $a$ and $b$
require a delay of $t$ time steps to meet for the first time after
time step $T$. Since these distributions will be the building blocks
that allow us to compute when a bundle can be delivered to its
destination, we call them {\em delivery distributions}.  $D_{ab}$ can
directly be derived from the contact profile $C_{ab}$:
\begin{multline}
D_{ab}(T,t) = C_{ab}(T+t) \prod_{i=T}^{T+t-1} \bigl(1-C_{ab}(i)\bigr) \\
  \forall a,b\in\nodeset, \;\forall T,t\in\mathbb{N}
\end{multline}

As before, the domain of these functions can be extended to
${\mathbb{R}^+}^2$.

\begin{definition}
The {\em delivery distribution set}, $\distset$, holds all the functions
${f: {\mathbb{R}^+}^2 \rightarrow \mathbb{R}^+}$ such that
${\forall T \in \mathbb{R}^+:} \; \int_0^\infty f(T,x) \,dx \leq 1$.
\label{def-D}
\end{definition}
Notice the inequality.

\begin{figure}
\begin{center} 
\includegraphics[scale=.75,bb=78 66 371 230]{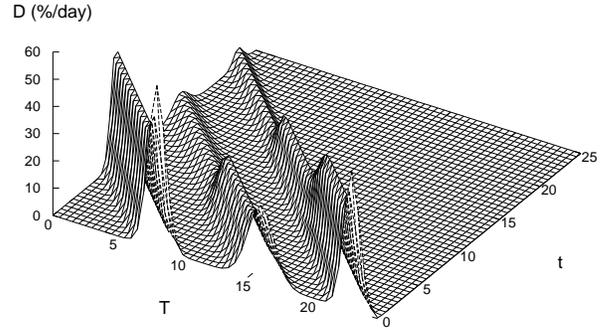}
\end{center}
\caption{\textbf{The \boldmath$D_{ab}(T,t)$ function} matching the
  contact profile given in figure~\ref{fig-contacts}.}
\label{fig-distrib}
\end{figure}

The $D_{ab}(T,t)$ distribution corresponding to the contact profile
given in figure~\ref{fig-contacts} is plotted in
figure~\ref{fig-distrib}.  Figure~\ref{fig-distrib9} plots the
function $D_{ab}(9,\cdot)$ ({\em i.e.} a section of $D_{ab}(T,t)$ in
the $T=9$ plane); the $D(T,\cdot)$ functions of course belong to
$\contactset$ (\mbox{$\forall T \geq 0$}).

Notice that $D_{ab}(T,\cdot)$ is the expected delivery delay
distribution for a bundle sent directly from a source $a$ to a
destination $b$ if $a$ decides to send it at time~$T$.

\begin{figure*}[htb]
\begin{center}
\includegraphics[scale=0.7]{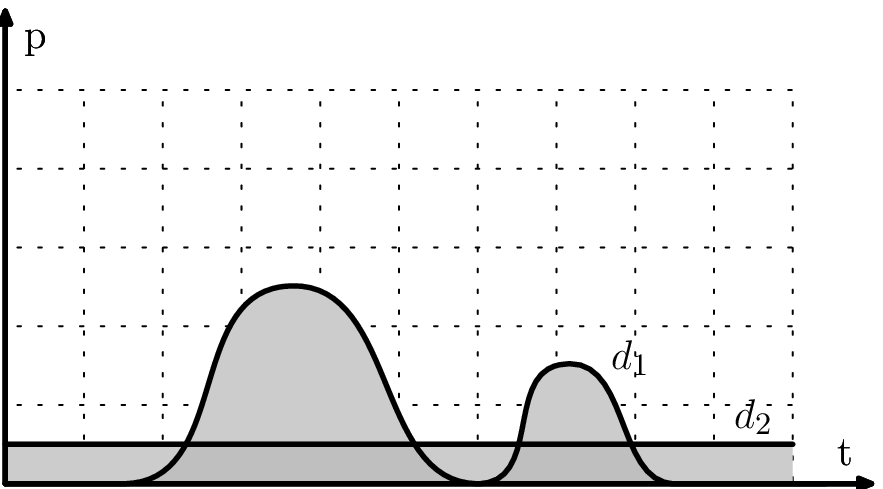}\hfill
\includegraphics[scale=0.7]{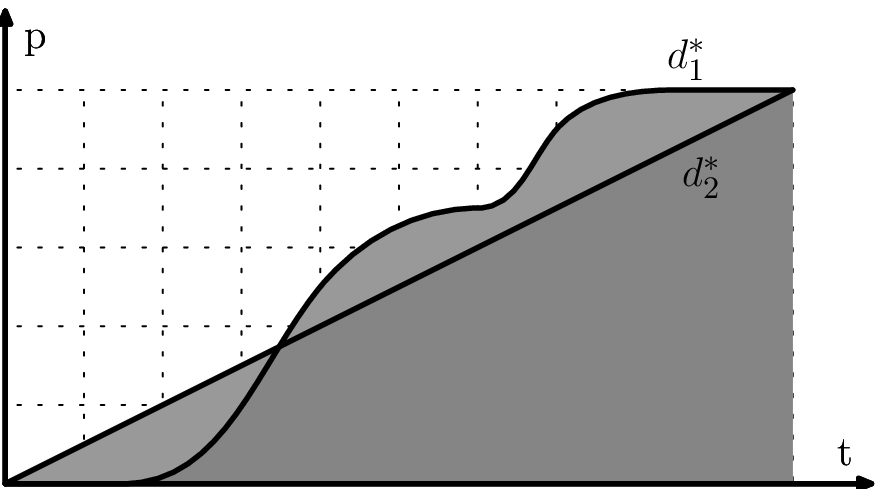}\hfill
\begin{minipage}[b]{5cm}
Definition~\ref{def-order} specifies when two contact distributions
$d_1,d_2$ are such that $d_1 \succeq d_2$.  The plots show two
distribution examples (left-hand plot) and their cumulative function
(denoted $d_1^*$ and $d_2^*$, right-hand plot).  We have $d_1 \succeq
d_2$ iff $\forall t\geq0: d_1^*(t) \geq d_2^*(t)$. Here, neither $d_1
\succeq d_2$ nor $d_2 \succeq d_1$ hold.
\end{minipage}\\[3mm]
\begin{minipage}{7cm}
\includegraphics[scale=0.7]{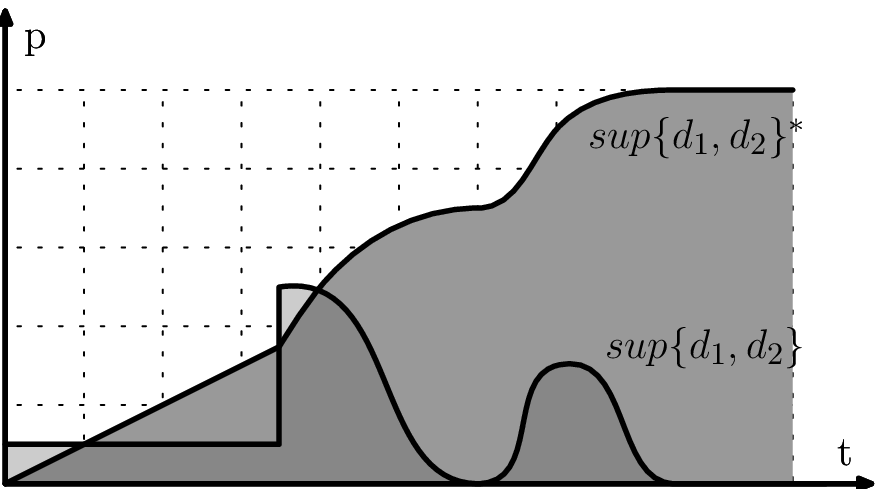}
\end{minipage}\hfill
\begin{minipage}{11cm}
The distribution $sup\{d_1,d_2\}$ (left-hand plot) is called {\em
supremum} (or {\em least upper bound}).  Its cumulative function
is the maximum of the $d^*_1$ and $d^*_2$ functions;
its distribution is the derivative of the cumulative function.

By definition of $sup\{d_1,d_2\}$, if ${d\succeq d_1}$ and ${d\succeq
d_2}$, then ${d\succeq sup\{d_1,d_2\}}\; {(\forall d\in\contactset)}$.
The infimum (or {\em greatest lower bound}) is defined in a similar
manner.

Since every element of $\contactset^2$ has a corresponding supremum
and infimum, the $\succeq$ relation defines a {\em lattice}
structure on $\contactset$ (and on $\distset$).
\end{minipage}
\end{center}
\caption{The $\succeq$ relation: example.}
\label{fig-order}
\end{figure*}

\subsection{Order relation on distributions}

We define an order relation between first contact distributions.
Intuitively, this relation allows us to compare two distributions to find
which one represents more frequent or predictable contacts.  A rigorous
definition is given below.

\begin{definition}
\label{def-order}
The first contact distributions $d_1 \in \contactset$ is {\em greater
(or equal)} than $d_2 \in {\cal C}$ (denoted $d_1 \succeq d_2$) if and
only if:
\begin{equation}
\forall x \geq 0: \quad \int_0^x d_1(t) \,dt \geq \int_0^x d_2(t) \,dt
\end{equation}
\label{eq-def-order}
\end{definition}

This relation is a {\em partial} order (but not a total order as there
exist $d_1, d_2\in\contactset$ such that neither ${d_1 \succeq d_2}$
nor ${d_1 \preceq d_2}$).  Figure~\ref{fig-order} gives an example of
incomparable first contact distributions.

It appears difficult to define a total order on
$\contactset$: comparing the distributions $d_1$ and $d_2$ in
figure~\ref{fig-order} is a matter of choice and depends on the
bundle delivery guarantees one wants to enforce.  The $\succeq$
relation is thus a least common denominator, and could be replaced in
what follows with a more restrictive order definition.

The worst (smallest) element of $\contactset$ is the $\bot$ ({\em
bottom}) distribution: $\bot(t) = 0$ ($\forall t\geq0$).  The best
(greatest) first contact distribution is denoted $\top$ ({\em top}):
$\top(t) = \delta(t)$ ($\forall t\geq0$); the $\delta$ symbol denotes
the Dirac distribution.

The $\succeq$ relation can be extended to $\distset$. For all
${D_1,D_2\in\distset}$:
$$
D_1 \succeq D_2 \iff 
  \forall T \geq 0: \; D_1(T,\cdot) \succeq D_2(T,\cdot)
$$ 

The $D_\bot$ delivery distribution is such that ${\forall T\geq0}:
{D_\bot(T,\cdot)\equiv\bot}$.  The definition of $D_\top$ follows immediately.

%
\section{Delivery distribution operators}
\label{sec-operators}
%

\subsection{The {\em forwarding} operator}

Let $D_{sbd}$ be the delivery distribution associated with the
delivery of a bundle from a source node $s$ to a destination $d$
via node $b$.  More precisely, if $s$ decides to send a bundle at
time~$T$, it will reach $d$ after a delay described by the
$D_{sbd}(T,\cdot)$ distribution. $D_{sbd}$ can be computed thanks to
$D_{sb}$ and $D_{bd}$:
\begin{equation}
D_{sbd} \equiv D_{sb} \otimes D_{bd}
\label{eq-def-times}
\end{equation}

The $\otimes$ (or {\em forwarding}) operator is a function defined for all
distribution pair.  We have $\otimes:\distset^2\rightarrow\distset$:
\begin{equation}
\bigl(D_1 \otimes D_2\bigr)(T,t) =  
   \int_0^t D_1(T,x) \, D_2(T+x,t-x) \,dx
\label{eq-def-times-proba}
\end{equation}

It is easy to see that this operator is associative but not
commutative.

Equation~\eqref{eq-def-times-proba} simply states that since the total
delivery delay is equal to $t$, if the delay to reach $b$ is equal to
$x$, then the delay from $b$ to $d$ is $t-x$.

Equation~\eqref{eq-def-times} can be generalized: a bundle could be
forwarded through several intermediate hops before reaching its
destination.  We denote $D_{s-d}$ (notice the dash) the delivery delay
distribution for a bundle sent from a source $s$ to a destination $d$
at time $T$; from now on, $\otimes$ will thus be applied to any
kind of delivery distributions.

For example, the graph below depicts a simple {\em delivery path},
{\em i.e.} a sequence of forwarding nodes; the corresponding delivery
distribution is also given.
\vskip2mm
$\xymatrix@C=15pt  
{
s\ar@{->}[r] &  a\ar@{->}[r]   & b\ar@{->}[r]   & d
}:\quad
D_{s-d} \equiv D_{sa}\otimes D_{ab}\otimes D_{bd}$
\vskip2mm

We say that two delivery paths with a common source $s$ and
destination $d$ are {\em disjoint} if the intersection of the set of
nodes they involve is $\{s,d\}$.

\subsection{The {\em duplication} operator}
\label{sec-duplication}

\newcommand{\Ddup}[0]{D_{\raisebox{5pt}[5pt]{\xymatrix@1@R=-1pt@C=4pt
{ {s}\ar@{-}[r]\ar@{-}[dr] & d \\ & d \\ } } }}

Let \def\objectstyle{\scriptstyle} $\Ddup$ be the delivery
distribution associated with the delivery of a bundle from $s$ to $d$
if it is duplicated so as to follow the disjoint delivery paths
described by the distributions $D_{s-d}$ and $D'_{s-d}$.  We have:
\begin{equation}
\Ddup \equiv D_{s-d} \oplus D'_{s-d} 
\label{eq-def-plus}
\end{equation}

The $\oplus$ (or {\em duplication}) operator is a function
$\oplus:\distset^2\rightarrow\distset$, defined as follows:
\begin{multline}
\bigl(D_1 \oplus D_2\bigr)(T,t) =
   \left( 1-\int_0^t D_1(T,x)\,dx \right) D_2(T,t) + \\
 \hspace{-0.5cm}\left( 1-\int_0^t D_2(T,x)\,dx \right) D_1(T,t) 
\label{eq-def-plus-proba}
\end{multline}

The expected delay computed is that of {\em the first} bundle to reach
the destination $d$.  It is easy to see that $\oplus$ is associative
and commutative.  We decide that $\otimes$ has a higher precedence
than $\oplus$.

Equation~\eqref{eq-def-plus-proba} is the sum of two terms.  Each term
is the probability that the bundle reaches the destination after a
delay $t$ using one path and that the bundle following the other path
is not arrived yet.

Notice that we have both $D_1\oplus D_2\succeq D_1$ and $D_1\oplus
D_2\succeq D_2$ (appendix, corollary~\ref{corollary-plus_order}).
This means that, contrary to what happens in deterministic networks,
duplicating a bundle to send it along two paths can improve
performance: it is not the case that the best path always delivers
the bundle first.

The definition of this operator allows us to apply it to arbitrary
independent distributions (for example, involving duplication and
forwarding).  This allows the computation of the distribution
associated with a non trivial way to deliver a bundle, such as the one
depicted below; the corresponding distribution formula is given on the
right.  Two arrows leaving a node depict a duplication.

\vskip2mm
\noindent
\begin{minipage}{0.3\columnwidth}
$\def\objectstyle{}
\xymatrix@C=12pt@R=3pt 
{
                        &                         & e\ar@{->}[r]    & d    \\
s\ar@{->}[r]\ar@{->}[dr]& b\ar@{->}[r]\ar@{->}[ur]& f\ar@{->}[r]    & d    \\
                        & c \ar@{->}[rr]          &                 & d   \\
}
$
\end{minipage}
~:
\begin{minipage}{0.65\columnwidth}
\begin{multline*}
D_{s-d}\equiv\bigl(D_{sc} \otimes D_{cd}\bigr) \oplus\bigl(D_{sb} \,\otimes \\ 
 (D_{be} \otimes D_{ed} \oplus D_{bf} \otimes D_{fd})\bigr)
\end{multline*}
\end{minipage}
\vskip2mm

\begin{figure*}[htp]
\begin{center}
\includegraphics[scale=0.55]{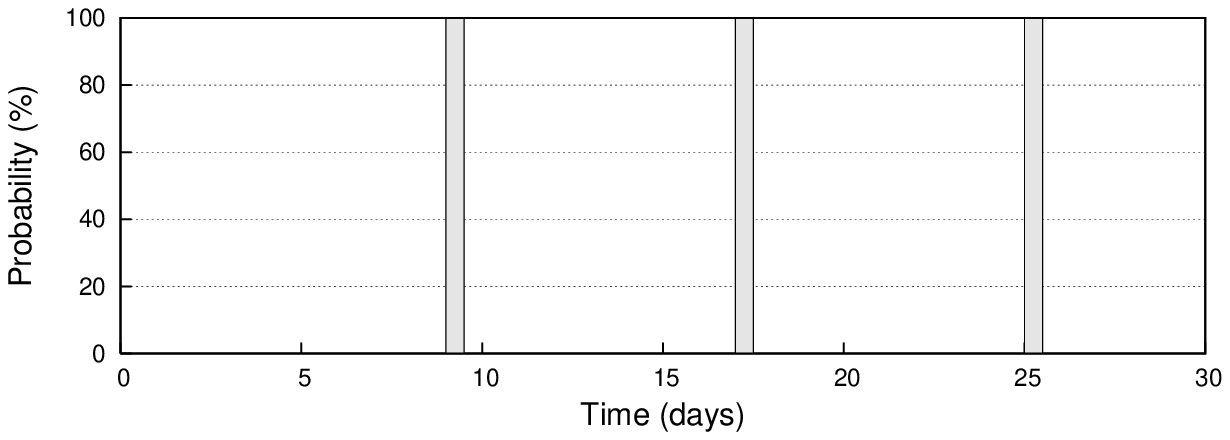} \hfill
\includegraphics[scale=0.5]{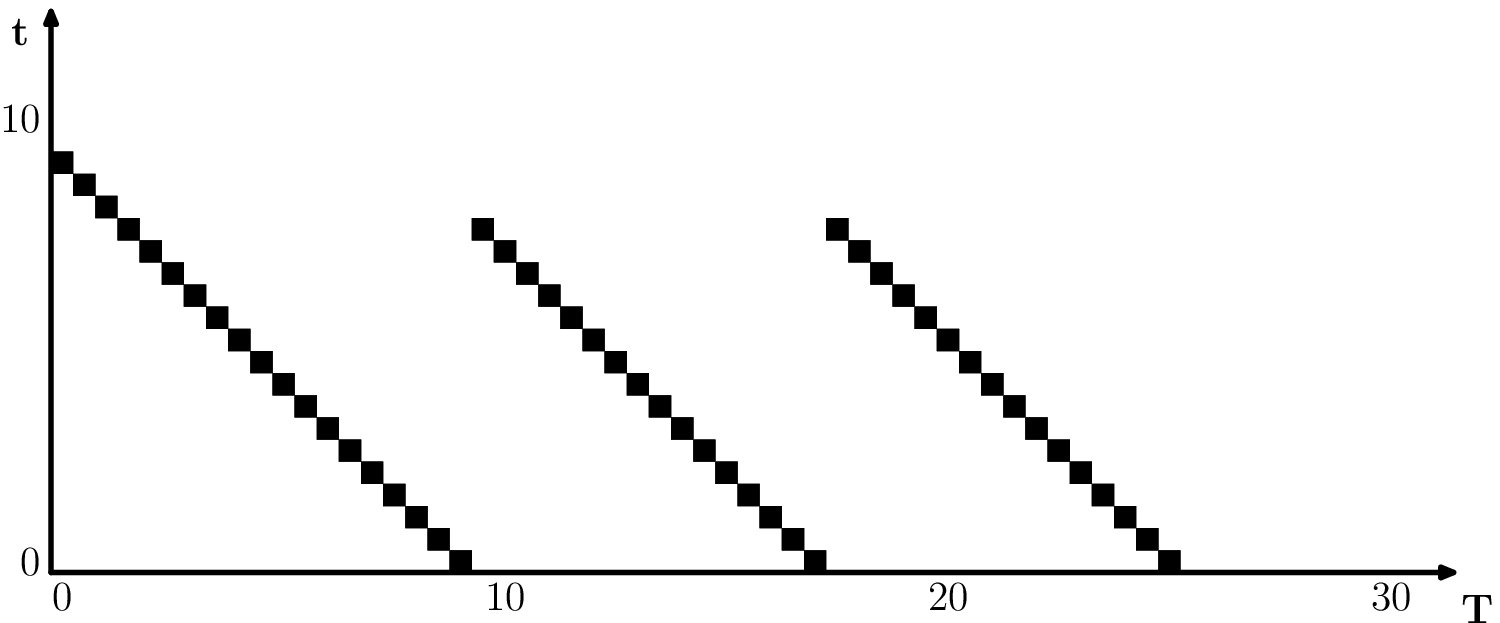} \\
\includegraphics[scale=0.7]{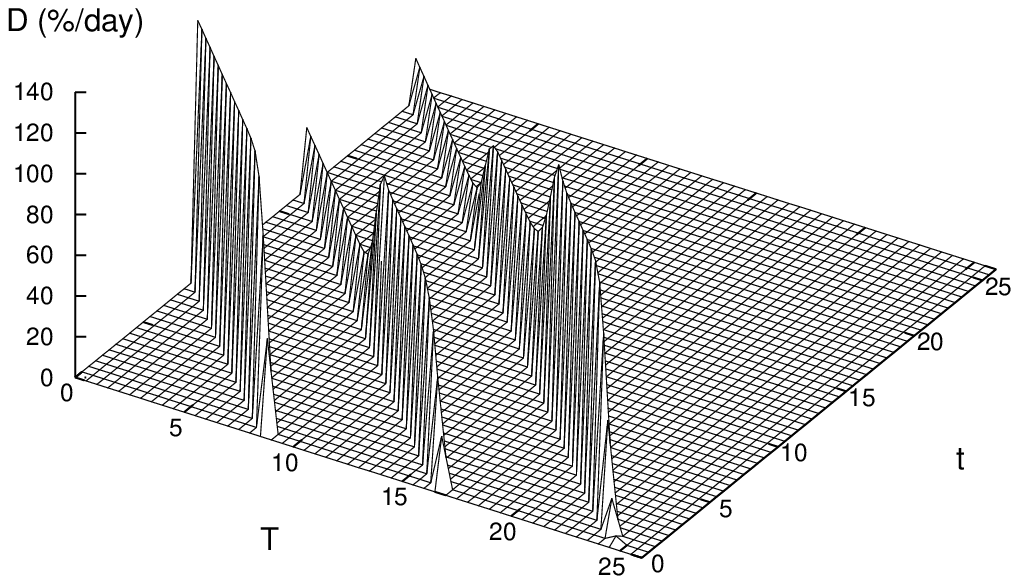}
\includegraphics[scale=0.7]{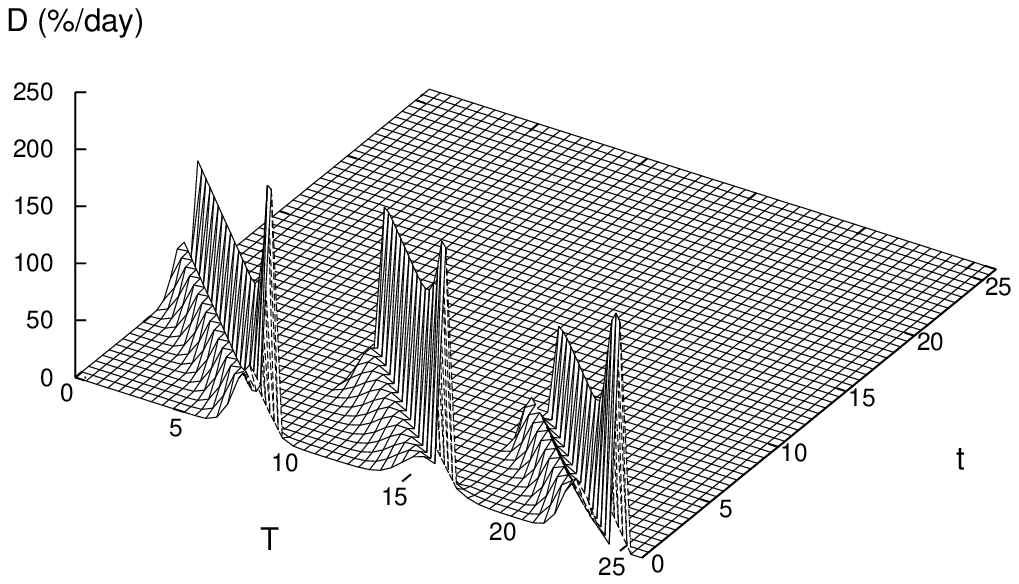}
\end{center}
\caption{\textbf{{\em Forwarding} ($\otimes$) and {\em duplication}
  ($\oplus$) operators: example}.  We denote $D_1$ the delivery
  distribution depicted in figure~\ref{fig-distrib}.  The top part of
  this figure depicts a contact profile (left) and the associated
  delivery distribution $D_2$ (dark squares represent a probability
  equal to 1).  The left 3D plot depicts $D_1 \otimes D_2$, the right
  one $D_1 \oplus D_2$.}
\label{fig-operators}
\end{figure*}

Figure~\ref{fig-operators} shows examples of the distributions
obtained using those operators.  As expected, the ``duplication''
operator shortens the delays and increases the delivery probability.

\subsection{The {\em scheduling} operator}

\newcommand{\Dsched}[0]{D_{\raisebox{5pt}[5pt]{\xymatrix@1@R=-1pt@C=4pt
{
{s}\ar@{-}[r]\ar@{.}[dr]     &  d  \\
                             &  d  \\
}}}                    }

Let \def\objectstyle{\scriptstyle} $\Dsched$ be the delivery
distribution that, every time a bundle has to be sent, chooses the
best delivery strategy out of $D_{s-d}$ and $D'_{s-d}$.  We have:
\begin{equation}
\Dsched \equiv D_{s-d} \schedule D'_{s-d}
\label{eq-def-schedule}
\end{equation}

The definition of $\schedule$ is straightforward. It is a function
${\schedule:\, \distset^2\rightarrow\distset}$ such that:
\begin{equation}
\bigl(D_1 \schedule D_2\bigr)(T,t) =
\begin{cases}
   D_1(T,t) &  \text{if $D_2(T,\cdot) \not\succeq D_1(T,\cdot)$} \\
   D_2(T,t) &  \text{otherwise}
\end{cases}
\label{eq-def-schedule-proba} \\
\end{equation}


If $s$ sends a bundle at time $T$, it is delivered using
$D_2(T,\cdot)$ if and only if $D_2(T,\cdot)\succeq D_1(T,\cdot)$.
This operator is not commutative since $\succeq$ is not a total order:
when $D_1(T,\cdot)$ and $D_2(T,\cdot)$ cannot be compared,
$D_1(T,\cdot)$ is chosen.  We decide that $\schedule$ has a lower
precedence than both $\otimes$ and $\oplus$.

The following example involves all the operators defined above.  Two
arrows leaving a node, one of them dotted, depict a scheduling
operation.  The dotted arrow leads to the second argument of
$\schedule$, emphasizing the operator's non-commutativity.

\vskip2mm
\noindent
\begin{minipage}{0.3\columnwidth}
$\def\objectstyle{}
\xymatrix@C=12pt@R=3pt 
{
                        &                         & e\ar@{->}[r]    & d    \\
s\ar@{->}[r]\ar@{->}[dr]& b\ar@{->}[r]\ar@{.>}[ur]& f\ar@{->}[r]    & d    \\
                        & c \ar@{->}[rr]          &                 & d   \\
}
$
\end{minipage}
~:
\begin{minipage}{0.65\columnwidth}
\begin{multline*}
D_{s-d}\equiv\bigl(D_{sc} \otimes D_{cd}\bigr) \oplus\bigl(D_{sb} \,\otimes \\ 
 (D_{bf} \otimes D_{fd}\schedule D_{be} \otimes D_{ed})\bigr)
\end{multline*}
\end{minipage}

\subsection{Delivery schemes}

We have defined a {\em delivery path} as a delivery strategy that only
involves forwarding.

A {\em delivery scheme} with source $s$ and destination $d$ is a
general delivery strategy that allows a bundle to be delivered from
$s$ to $d$.  It can use an arbitrary number of forwarding, duplication
and scheduling operations. A delivery path is thus a particular
delivery scheme.

Two delivery schemes from $s$ to $d$ are {\em disjoint} if the
intersection of the set of nodes they involve is $\{s,d\}$.

%
\section{Delivery guarantees}
\label{sec-guarantees}
%

Knowing the delay distribution $d_{s-d}\in\contactset$ associated with
the delivery of a bundle allows us to verify a large range of
conditions on permissible delays or on delivery probabilities.

For example, the condition 
$$\int_0^\infty d_{s-d}(t)\,t \,dt \leq d_\text{max}$$ 
imposes a maximum expected delay $d_\text{max}$, while 
$$
\int_0^\text{1h} d_{s-d}(t) \,dt \geq .9 \quad\text{and}\quad
\int_0^\text{24h} d_{s-d}(t) \,dt \geq .99
$$
matches distributions delivering a bundle in less than one hour nine
times out of ten, and in less than a day with a probability of 99\%.

We naturally impose that a condition fulfilled for a certain delivery scheme
must be fulfilled for better schemes.

\begin{definition}
\label{def-condition}
A {\em delivery condition} $C$ is a predicate:
\mbox{$C: \contactset \rightarrow \{\mbox{true}, \mbox{false}\}$} with
$\forall d_1,d_2\in\contactset \text{~such that~} d_1 \succeq d_2:
 C(d_2)\Longrightarrow C(d_1)$.
\end{definition}

A condition $C$ can be extended to a delivery distribution
$D\in\distset$: $C(D) \iff \forall T\geq 0:
C\bigl(D(T,\cdot)\bigr)$.

%
\section{Delivering bundles with guarantees}
\label{sec-algo}
%

\subsection{Probabilistic Bellman-Ford}

Algorithm~\ref{algo-fast} adapts the Bellman-Ford algorithm to
predictable disruption tolerant networks.  In this section, {\em we do not
allow bundle duplication}.  Notice that, in general, the concept of
``shortest path'' is meaningless since the $\preceq$ relation is a
{\em partial} order.

\begin{figure}
\restylealgo{ruled}
\begin{algorithm}[H]
\SetKwData{stabilized}{stabilized}

\KwData{$d$ is the destination node}
\BlankLine

$\forall\; x \in {\cal N}\setminus\{d\} :\; B_x \leftarrow D_\bot$\;
$B_d \leftarrow D_\top$\;
\BlankLine

\Repeat{\stabilized}{
\stabilized $\leftarrow$ true\;
\ForAll{\nllabel{algo-fast_for1}
          $x \in {\cal N}$}{
    \ForAll{\nllabel{algo-fast_for2}
              $y \in {\cal N}$}{
        \nllabel{algo-fast_via_y}
          $D_{xy-d} \leftarrow D_{xy} \otimes B_y$\;
        \nllabel{algo-fast_schedule}
          \If{$B_x \neq B_x \schedule D_{xy-d}$}{
        \stabilized $\leftarrow$ false\;
        \nllabel{algo-Bx_update}
          $B_x \leftarrow B_x \schedule D_{xy-d}$\;
    }
  }
}
}
\caption{Probabilistic Bellman-Ford}
\label{algo-fast}
\end{algorithm}
\end{figure}

Similarly to the Bellman-Ford algorithm, algorithm~\ref{algo-fast}
computes, for every node $n\in\nodeset$, the best distribution leading
to the destinations found so far ($B_n$).  This distribution is
propagated to its neighbours ({\em i.e.} all the other nodes since the
network is infrastructureless).

Once node $x$ receives the best delivery distribution $B_y$ found by
$y$, it computes the delivery distribution obtained if it would send
the bundle directly to $y$, and if $y$ would forward it according to
$B_y$.  The resulting distribution is denoted $D_{xy-d}$
(line~\ref{algo-fast_via_y}).

$D_{xy-d}$ is compared to the best known distribution to the
destination ($B_x$) by means of the $\schedule$ operator.  If
$D_{xy-d}$ is better than $B_x$ on some time intervals, $B_x$ is
updated (line~\ref{algo-Bx_update}).

The algorithm terminates once no more $B_x$ distribution is updated.

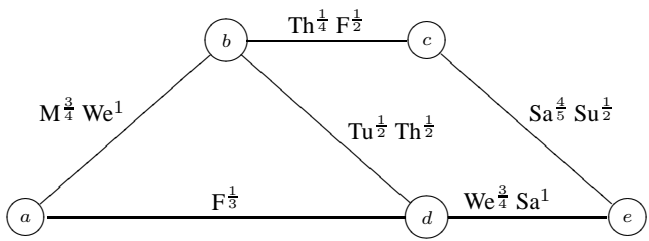
\begin{figure*}[htp]
\begin{minipage}{0.45\textwidth}
\xymatrix@C=60pt@R=50pt 
{
  &  *++[o][F-]{b}\ar@{-}[r]^{\txt{\small Th}\frac{1}{4} \;
                              \txt{\small F}\frac{1}{2}}
                  \ar@{-}[dr]^(.6){\txt{\small Tu}\frac{1}{2} \;
                               \txt{\small Th}\frac{1}{2}} 
      & *++[o][F-]{c}\ar@{-}[dr]^{\txt{\small Sa}\frac{4}{5} \;
                                  \txt{\small Su}\frac{1}{2}}
           &         \\
*++[o][F-]{a}\ar@{-}[rr]^{\txt{\small F}\frac{1}{3}}
             \ar@{-}[ur]^{\txt{\small M}\frac{3}{4} \;
                          \txt{\small We}1} 
  &
      & *++[o][F-]{d}\ar@{-}[r]^(.4){\txt{\small We}\frac{3}{4} \;
                                 \txt{\small Sa}1}
           & *++[o][F-]{e} \\
}
\end{minipage}
\hfill
\begin{minipage}{0.50\textwidth}
This graph gives the contact profiles of the nodes
$a,b,c,d,{e\in\nodeset}$.

Unconnected nodes never meet each other: they have a null contact
profile (and a corresponding delivery distribution $D_\bot$).

The label connecting the other nodes describes which days they might
have a contact. For example, there is one chance out of four that $b$ and
$c$ meet on Thursday, and one out of two on Friday.
\end{minipage}
\caption{\textbf{A predictable network.}}
\label{fig-algo_ex_contacts}
\end{figure*}

\begin{figure*}[htp]
\begin{minipage}{0.45\textwidth}
The opposite table shows how our probabilistic Bellman-Ford algorithm
behaves.  This example is based on a simple network made of 5 nodes.
The nodes contact profiles are given in
figure~\ref{fig-algo_ex_contacts}.  In this example, $a$ is the source
node and $e$ is the destination.

At first, all the nodes (but the destination) have no knowledge of any
path to the destination; their best distribution is thus set to
$D_\bot$.  The destination's delivery distribution to itself is of
course $D_\top$.

Line~2 depicts the results obtained after the first iteration.  Since
only $c$ and $d$ have contacts with the destination, only $B_c$ and
$B_d$ are modified.  They are set to the direct contact with the
destination distribution since, for example, ${D_{c-e} \otimes D_\top =}
D_{c-e}$.  The delivery distributions are depicted as a square plot;
the $x$-axis is the bundle sending time, the $y$-axis is the delay to
reach the destination.  Each square represents a 24 hour period, the
first column matches bundles sent on Monday.

During the next iteration (line~3), $a$ discovers it might meet with
$d$ before $d$ meets $e$.  $B_a$ is thus changed to ${D_{a-d} \otimes
D_{d-e}}$.  The bundles received by $b$ can be forwarded to $c$ or
$d$.  The distributions $D_{b-c-e}$ and $D_{b-d-e}$ are thus compared;
bundles sent Tuesday or before are sent {\em via} $d$, those sent after
Tuesday are sent {\em via} $c$.

The last iteration allows $a$ to decide when bundles should be sent to
$b$ or $d$.  The distributions $B_a$ and $D_{a-b}\otimes B_b$ are thus
compared; the latter is given between parentheses.  Neither $c$ nor
$d$ should forward bundles to $b$, thus $B_c$ and $B_d$ are left
untouched.

The algorithm is stabilized since neither $b$, $c$, or $d$ should
forward bundles {\em via} $a$.
\end{minipage}
\hfill\begin{tabular*}{0.5\textwidth}[]{cl}
\bf 1 & $B_a\equiv B_b\equiv B_c\equiv B_d\equiv D_\bot
        \quad B_e\equiv D_\top$
        \vspace*{.3cm}\\
\hline\\
\bf 2 & $\vcenter{\hbox{\hspace*{0.3cm}$B_c\equiv D_{c-e}$}
                  \hbox{\includegraphics[scale=0.40]{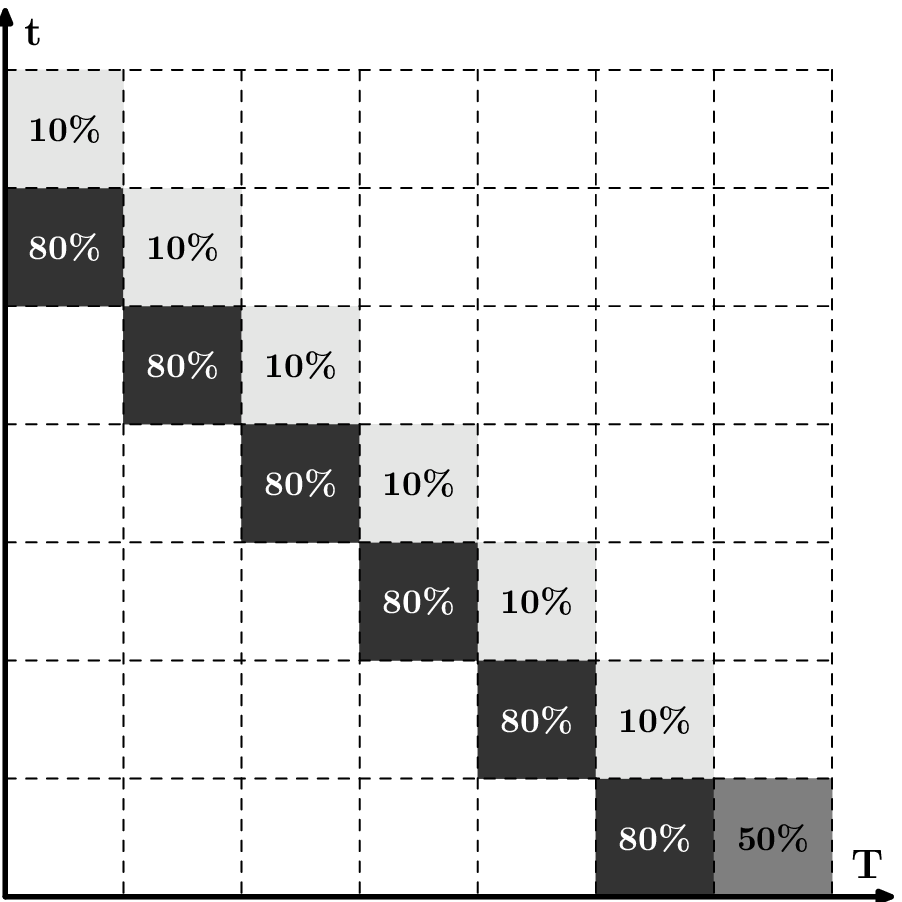}}}$
        \hspace*{.5cm}
        $\vcenter{\hbox{\hspace*{0.3cm}$B_d\equiv D_{d-e}$}
                  \hbox{\includegraphics[scale=0.40]{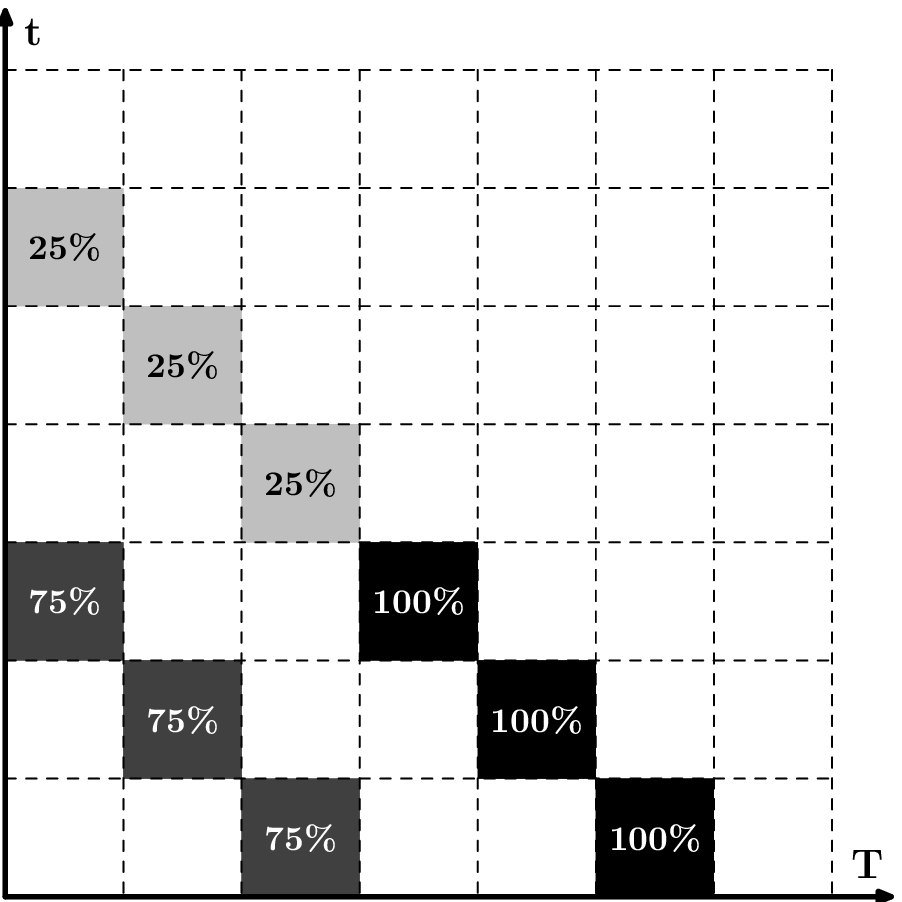}}}$
        \vspace*{.3cm}\\
\hline\\
\bf 3 & $\vcenter{\hbox{\hspace*{0.3cm}$B_a\equiv D_{a-d-e}$}
                  \hbox{\includegraphics[scale=0.40]{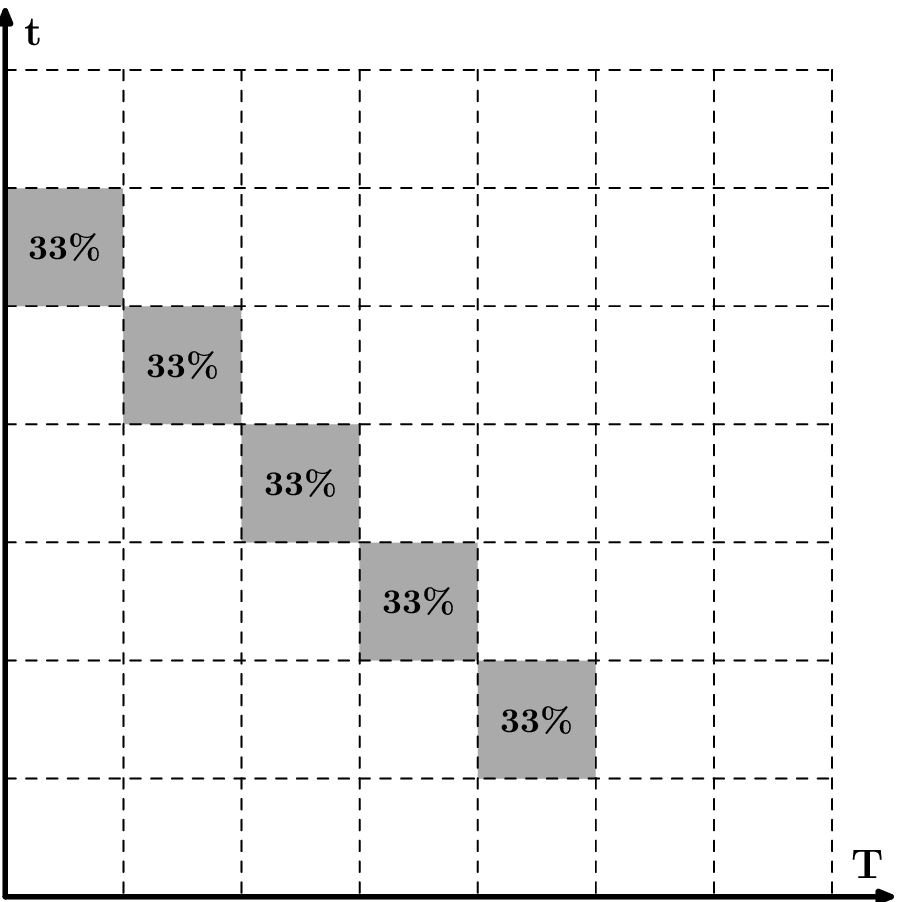}}}$
        \hspace*{.5cm}
        $\vcenter{\hbox{\hspace*{0.3cm}$B_b$}
                  \hbox{\includegraphics[scale=0.40]{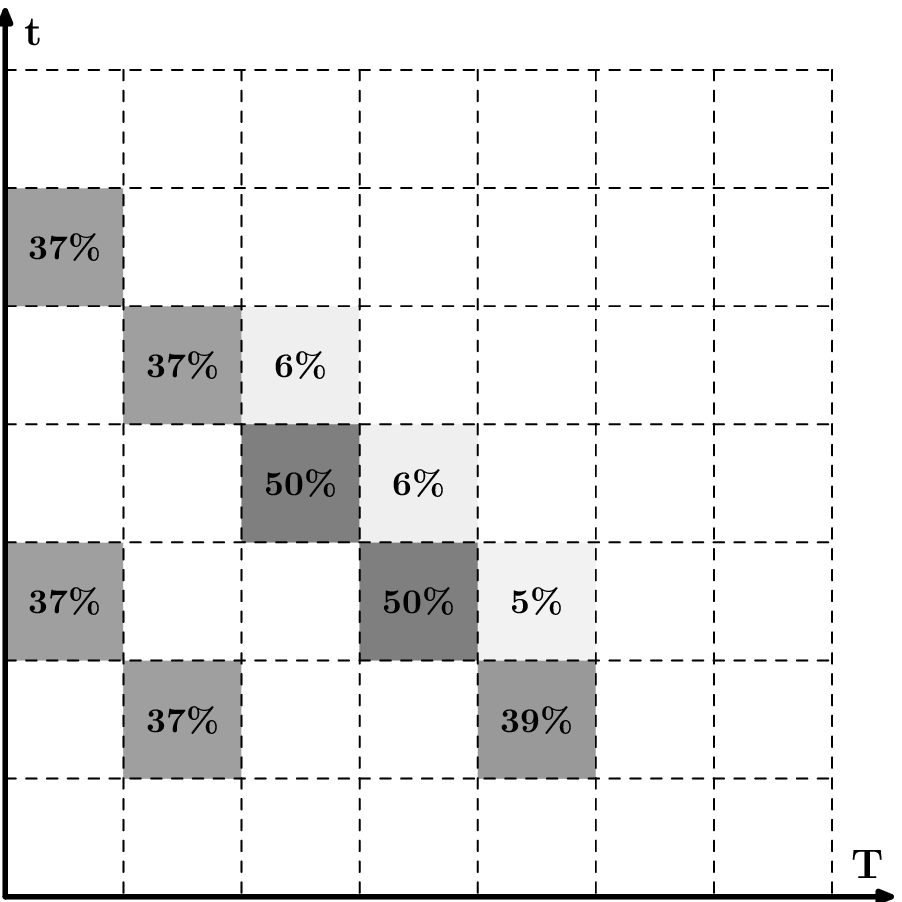}}}$
        \vspace*{.3cm}\\
\hline\\
\bf 4 & $\vcenter{\hbox{\hspace*{0.3cm}$B_a$}
                  \hbox{\includegraphics[scale=0.40]{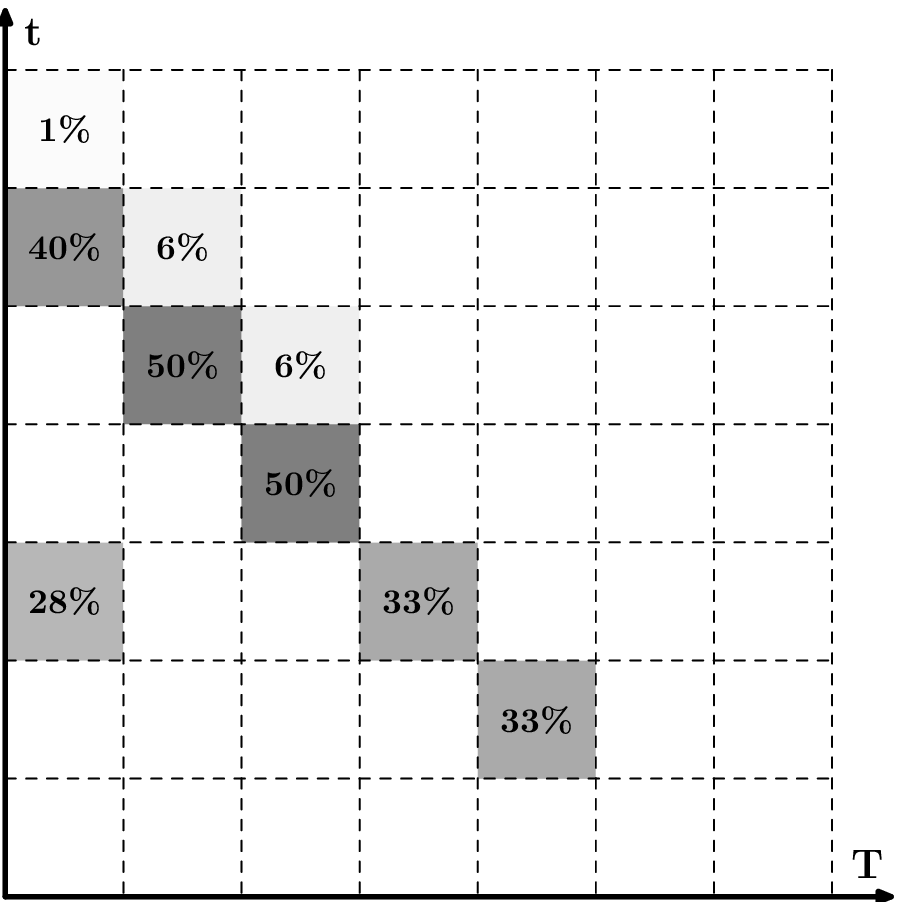}}
        \vspace*{.3cm}}$
        \hspace*{.5cm}
        $\left(\vcenter{\hbox{\hspace*{0.3cm}$D_{a-b}\otimes B_b$}
                  \hbox{\includegraphics[scale=0.40]{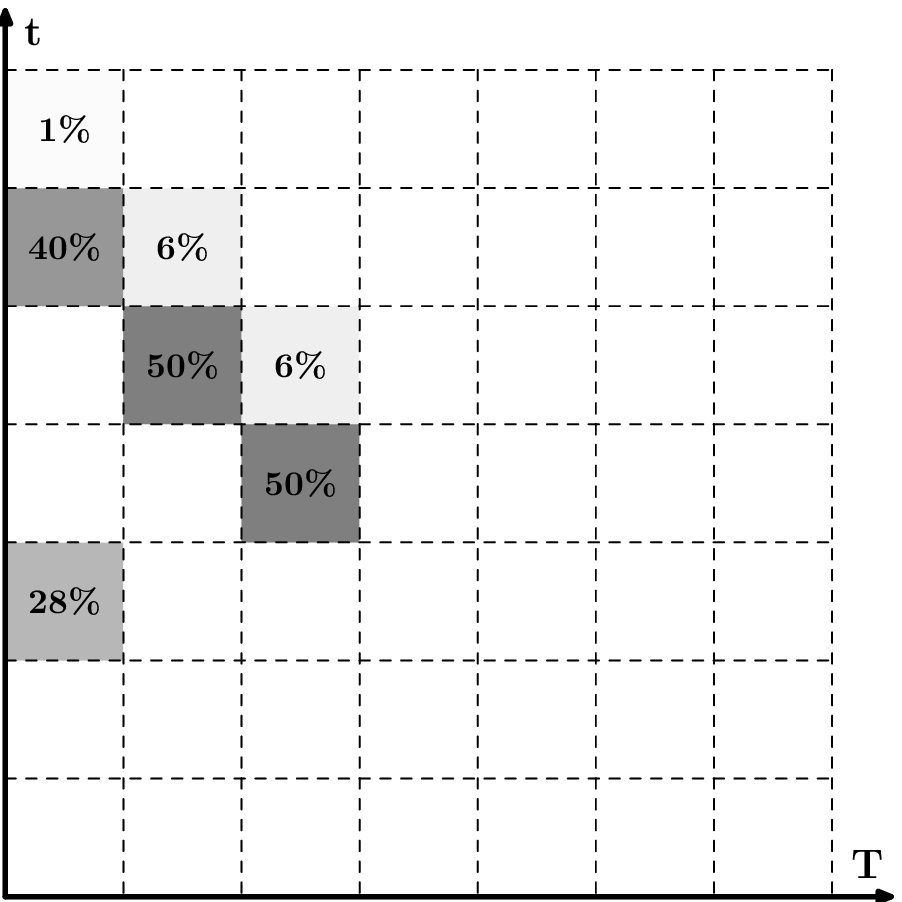}}
        \vspace*{.3cm}}\right)$
\end{tabular*}
\caption{\textbf{Probabilistic Bellman-Ford:} example.}
\label{fig-algo_ex_main}
\end{figure*}

Figures~\ref{fig-algo_ex_contacts} and~\ref{fig-algo_ex_main}
demonstrate how the algorithm works by means of a small example.

As mentioned before, this algorithm generalizes both \cite{TZZ03}
({\em i.e.} converges to the {\em ``shortest expected path''}) and
\cite{M04}\footnote{To be fair, this work also deals with message
transmission delays, which are not considered here.} ({\em i.e.} finds
the exact shortest path in the case of perfectly predictable
networks).

The delivery computed by this algorithm depends on the order at which
the elements of $\nodeset$ are picked up (lines \ref{algo-fast_for1}
and \ref{algo-fast_for2}).  In practice, it might be preferable to
rely on a heuristic to choose the preferred elements first.

\subsection{Guarantees}

Our aim is now to find a way to deliver bundles that fulfills a given
condition $C$ as specified in definition~\ref{def-condition}, while
trying to minimize the network's bandwidth/energy/memory consumption.

Ideally, the DTN is predictable enough to enforce condition~$C$
without duplicating any bundle.  We thus propose to rely on
algorithm~\ref{algo-fast} to find a first delivery scheme (and,
thus, a first delivery distribution $D_1$).

If $C$ is not fulfilled by $D_1$, we search for another
fast bundle forwarding scheme using algorithm~\ref{algo-fast}; let
$D_2$ be its delivery distribution.  We then duplicate the bundle on
both delivery schemes, yielding a distribution $D_1\oplus D_2$.  We
have already pointed out that $D_1\oplus D_2 \succeq D_1$, thus
$C(D_1\oplus D_2)$ is more likely to be {\em true} then $C(D_1)$.

This process is iterated until $C$ is finally fulfilled.

As mentioned in section~\ref{sec-duplication}, the distribution
computed by the ``duplication'' ($\oplus$) operator is biased if its
operands are not {\em independent} distributions.  The simple
distribution formula $(D_{sb}\otimes D_{bd}) \oplus (D_{sb}\otimes
D_{bd})$ brings to light the problem caused by dependent distributions.

To avoid this bias, we ensure that $D_1$ and $D_2$ are independent by
forbidding $D_2$ to rely on the nodes involved in $D_1$ (source and
destination nodes excluded, line~\ref{algo-constrained_proba_rm}).

The resulting algorithm is given below.

\vspace{2mm}
\dontprintsemicolon
\begin{algorithm}[H]
\KwData{Network nodes $\nodeset$; delivery condition $C$}
\KwData{Bundle source $s$ and destination $d$}
\SetKw{Or}{or}
\BlankLine

$B \leftarrow D_\bot$ \;
\BlankLine
\Repeat{$C(B)$ \Or $\nodeset=\{s,d\}$}{
  Using nodes in $\nodeset$, compute $D\in\distset$ {\em via}
    algorithm~\ref{algo-fast} \;
  $B \leftarrow B \oplus D$ \;
  $\nodeset \leftarrow \nodeset \setminus \{\text{nodes
  involved in $D$}\} \cup \{s,d\}$ 
  \nllabel{algo-constrained_proba_rm}\;
}
\caption{Constrained probabilistic delivery}
\label{algo-constrained_proba}
\end{algorithm}
\vspace{2mm}

Nothing guarantees of course that there exists a way to deliver bundles
that satisfies $C$: even an epidemic broadcasting might not suffice.

\subsection{More on disjoint delivery schemes}

The {\em constrained probabilistic delivery} algorithm above computes
a delivery scheme that consists of duplicating the bundle to multiple,
independent, non-duplicating delivery schemes.

To ensure independence, algorithm~\ref{algo-constrained_proba} enforces
those non-duplicating delivery schemes to operate on completely
distinct node sets.  This might be too stringent if the network is
small or sparse.  We thus propose to allow such a delivery scheme to
use nodes that are {\em unlikely} to receive a bundle according to the
other schemes.  The resulting delivery distributions will thus be {\em
almost} independent.

Line~\ref{algo-constrained_proba_rm} of
algorithm~\ref{algo-constrained_proba} is thus changed: only the nodes
involved in $D$ with a probability higher than a given threshold are
removed.  The specific value of this threshold is a function of the
network considered.

The rest of this section explains how to compute the
probability that a given node receives a bundle, given a
(non-duplicating) delivery scheme computed by
algorithm~\ref{algo-fast}.

We have seen that the proposed modified Bellman-Ford algorithm does
not lead to a simple routing table: if a bundle reaches a given node
at time $T$, its next hop depends on its destination {\em and on $T$}.
Each node $n$ divides time in intervals $I^n_1, I^n_2,\dots$ (by means
of the $\schedule$ operator, algorithm~\ref{algo-fast}
line~\ref{algo-fast_schedule}), and each interval $I$ matches a given
next hop $H_n(I)$.  In the example figure~\ref{fig-algo_ex_main},
node~$b$ has defined two intervals:
$I^b_1=[\mbox{Monday},\mbox{Tuesday}]$ and
$I^b_2=[\mbox{Wednesday},\mbox{Sunday}]$; $H_b(I^b_1)=d$ and
$H_b(I^b_2)=c$.

\begin{figure}[b!ht]
\begin{center}
\includegraphics[scale=1]{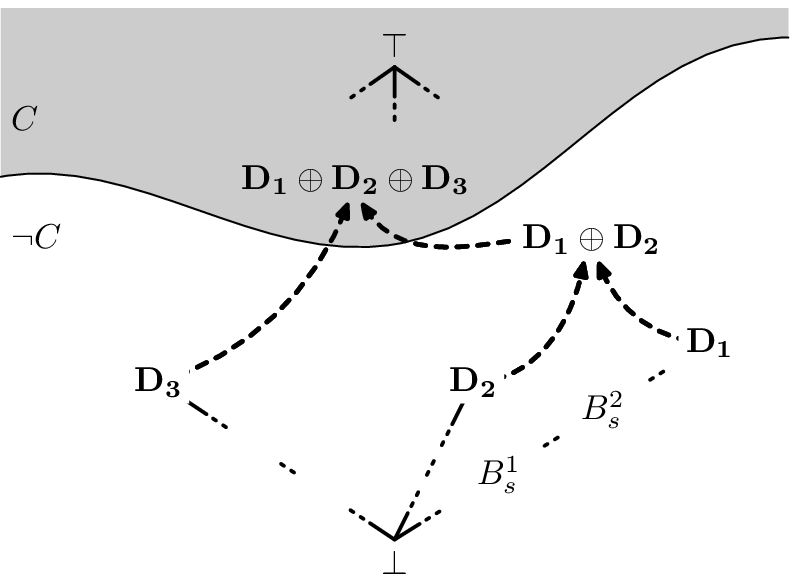}
\end{center}
\caption{\textbf{Finding a fast delivery scheme that fulfills
  condition $C$ using algorithms~\ref{algo-fast} and
  \ref{algo-constrained_proba}}.  This figure represents the delivery
  distribution lattice (introduced in figure~\ref{fig-order}); it is
  depicted the usual way (basically, an element is greater than
  another one if it is placed above). The greyed area corresponds to
  elements that satisfy condition $C$. \textbf{1.} The adapted
  Bellman-Ford algorithm is used to find a distribution ($D_1$) that
  characterizes a fast way to deliver bundles; $B_s^i$ denotes the
  source node's best distribution found after $i$ iterations. We have
  $\bot\preceq B^1_s\preceq B^2_s\preceq\dots\preceq D_1$. \textbf{2.}
  Since $\neg C(D_1)$, another disjoint delivery distribution, $D_2$,
  is computed using algorithm~\ref{algo-fast}.  Combined with $D_1$, it
  leads to $D_1\oplus D_2$ which still does not satisfy $C$. $D_3$ is
  thus computed, and combined with $D_1$ and $D_2$, gives a satisfactory
  delivery scheme $D_1\oplus D_2\oplus D_3$.  We have $D_1\preceq
  D_1\oplus D_2\preceq D_1\oplus D_2\oplus D_3$.}
\label{fig-algo}
\end{figure}

A bundle crosses a number of nodes on its way to its destination.
We compute the probability $P_n$ that a given node $n$ is one of them.

Let $s$ be the bundle source node and $d$ the destination.  If the
bundle is ready to be sent at time~$T$, it should reach $n=H_s(I)$,
where $I$ is the time interval of $s$ such that $T \in I$. The bundle
arrival time at $n$ follows the contact distribution ${N(x) =}
D_{sn}(T,x)$, thus ${P_n = \int_0^\infty N(x) \,dx}$.

Once the bundle has been received by $n$, each time interval $I^n_i$
matches a potential next hop $n_i=H_n(I^n_i)$.  We have:
\begin{eqnarray}
N^n_i(x) &=& \int_{\{t \in I^n_i \mid t \leq x\}} \!\!\!\!\!
   N(t) \,D_{nn_i}(t,x-t) \,dt 
   \label{eq-disjoint-hop-dist}\\
P_{n_i} &=& \int_0^\infty N^n_i(x) \,dx
   \label{eq-disjoint-hop-proba}
\end{eqnarray}

Equation~(\ref{eq-disjoint-hop-dist}) gives the bundle time arrival
distribution at the next hop~$n_i$.  This process can be continued
recursively until the probability of receiving the bundle is known for
all nodes.

The bundle forwarding process can be represented by a graph.  The
children of a node are the potential next hops.  
The graph obtained for the example depicted in
figure~\ref{fig-algo_ex_main} is given below.

\vskip2mm
\centerline{\xymatrix@C=20pt@R=10pt
{
  &  
      & *++[o][F-]{c} \ar @{} [dr] |(.4) {5/32}
                      \ar@{->}[r]
           &  *++[o][F-]{e} \ar @{} [r] |(.9) {25/256}
              &  \\
*++[o][F-]{a}\ar @{} [u] |(.5) 1
             \ar@{->}[r]
             \ar@{->}[dr]
  & *++[o][F-]{b} \ar @{} [u] |(.5) 1
    \ar@{->}[ur]
    \ar@{->}[r]
      & *++[o][F-]{d} \ar @{} [dr] |(.4) {9/16}
           \ar@{->}[r]
           & *++[o][F-]{e} \ar @{} [r] |(.7) {9/16}
               & \\
  &  *++[o][F-]{d}\ar @{} [ur] |(.3) 0
                  \ar@{->}[rr]
      &
           & *++[o][F-]{e} \ar @{} [r] |(.5) 0
               &
}}
\vskip3mm

The numbers labelling the nodes are the probabilities of receiving the
bundle, as given by (\ref{eq-disjoint-hop-proba}).  The destination
$e$ thus receives the bundle with a probability of
$\frac{25}{256}+\frac{9}{16}$.


%
\section{Conclusion and future works}
\label{sec-conclusions}
%

We propose to model contacts between a disruption tolerant network's mobile
nodes as a random process, characterized by contact distributions.
Such a description is more general than those generally encountered in
the literature, and allows, for example, to model a perfectly
deterministic network.

We show how such contact distributions can be combined to compute the
bundle delivery delay distribution corresponding to a given delivery
strategy ({\em i.e.} a description of the nodes forwarding
decisions).  We show how the Bellman-Ford algorithm can be adapted to
cope with such stochastic networks.  

There is a tradeoff between a bundle's delivery probability/delay and
the consumption of network resources.  We propose to duplicate
bundles along disjoint ``shortest'' path so as to meet a given delivery
guarantee without consuming too many resources.  The corresponding
algorithms are given.

This work can be continued along several lines.  

We have proposed a way to route bundles through the network; other
routing strategies should be explored and compared.

Three operators on delivery distributions have been defined.  Others
could be added so as to describe more subtle routing decisions, or to
deal with bundles' transmission delays.

Real network traces should be analysed so as to quantify their
predictability, to compare delivery strategies, and to measure how
predictability impacts performance.

\appendix
%
%


\begin{lemma}
\label{lemma-plus_order}
$\forall D_1, D_2, D_3 \in {\cal D}$, we have
$D_2 \succeq D_3 \Rightarrow D_1 \oplus D_2 \succeq D_1 \oplus D_3$.
\end{lemma}
\begin{proof}
Given the definition of $\oplus$ and $\succeq$, one must prove that,
$\forall D_1, D_2 \in {\cal D}, \; \forall T \geq 0, t \geq 0$,
given $D_2 \succeq D_3$:
\begin{multline}
\int_0^t \left[ 1-\int_0^x D_1(T,y)\,dy \right] D_2(T,x) +\\
  \left[ 1-\int_0^x D_2(T,y)\,dy \right] D_1(T,x) \,dx \\
\geq\int_0^t\left[ 1-\int_0^x D_1(T,y)\,dy \right] D_3(T,x)+\\
  \left[ 1-\int_0^x D_3(T,y)\,dy \right] D_1(T,x) \,dx
\label{eq-plus_order-first}
\end{multline}

The left-hand part can be written as:
\begin{multline}
\int_0^t D_1(T,x) \,dx + \int_0^t D_2(T,x) \,dx - \mbox{}\\
  \int_0^t \int_0^x \left[ D_1(T,x) D_2(T,y) + 
    D_1(T,y) D_2(T,x) \right] \,dy\,dx
\label{eq-plus_order-second}
\end{multline}

Changing the double integral's integration order, the last term of
\eqref{eq-plus_order-second} is equal to:
\begin{eqnarray}
& & \int_0^t D_1(T,x) \int_0^x D_2(T,y) \,dy \,dx + \nonumber \\
& &  \hspace*{10mm} \int_0^t  D_1(T,x) \int_x^t D_2(T,y) \,dy \,dx \nonumber \\
&=& 
\int_0^t D_1(T,x) \; dx \int_0^t D_2(T,y) \,dy
\end{eqnarray}

The same procedure can be applied to the right-hand part of
(\ref{eq-plus_order-first}). (\ref{eq-plus_order-first}) is thus equivalent to:
\begin{equation}
\int_0^t D_2(T,x) \,dx \geq \int_0^t D_3(T,x) \,dx
\end{equation}
Which holds by hypothesis.
\end{proof}

\begin{corollary}
\label{corollary-plus_order}
$\forall D_1, D_2 \in {\cal D}$, we have $D_1 \oplus D_2 \succeq D_1$.
\end{corollary}
\begin{proof}
From Lemma~\ref{lemma-plus_order},
$D_1 \oplus D_2 \succeq D_1 \oplus \bot = D_1 $.
\end{proof}



\bibliographystyle{IEEEtran}
\bibliography{paper}

\end{document}